\documentstyle[prl,aps,twocolumn,multicol,epsf]{revtex}
\tighten
\begin{document}
\twocolumn[\hsize\textwidth\columnwidth\hsize\csname
@twocolumnfalse\endcsname
\title{Strength of the null singularity inside black holes}
\author{Lior M. Burko}
\address{
Theoretical Astrophysics 130--33,
California Institute of Technology, Pasadena, California 91125}
\date{\today}

\maketitle

\begin{abstract}
We study analytically the Cauchy horizon singularity inside
spherically-symmetric charged black holes, coupled to a spherical
self-gravitating, minimally-coupled, massless scalar field.  We show that
all causal geodesics terminate at the Cauchy horizon at a null
singularity, which is weak 
according to the Tipler classification. The singularity is also
deformationally-weak in the sense of Ori. Our results are valid at arbitrary
points along the null singularity, in particular at late retarded times,
when non-linear effects are crucial.
\newline  
\newline
PACS number(s): 04.70.Bw, 04.20Dw
\end{abstract}

\vspace{3ex}
]

\section{Introduction}
The issue of spacetime singularities -- which are known 
to inevitably occur inside black holes under very
plausible assumptions \cite{hawking-ellis73} -- is an intriguing puzzle of
physics. The laws of physics, as we presently understand them (e.g.,
classical General Relativity), are presumably invalid at singularities.
Instead, some other theories (e.g., quantum gravity), as yet unknown, are
expected to take over from General Relativity and control the spacetime 
structure. The General Relativistic predictions are nevertheless of the
greatest importance, as they reveal the spacetime structure under extreme
conditions in the strong-field regime. 
Of particular interest is the possibility that there are two distinct ways
in which General Relativity can fail at different types of singularities;
For one type of singularities the failure 
is through infinite destructive effects on physical objects, whereas for
the other type the failure is through breakdown of predictability.   

Until recently, the only known generic singularity in General relativity
was the Belinsky-Khalatnikov-Lifshitz (BKL) singularity \cite{bkl-70}.
According to the BKL picture, spacetime develops a succession of Kasner
epochs in which the axes of contraction and expansion change directions
chaotically. This succession ends at unbounded oscillations at a spacelike
singularity, which is unavoidably destructive for any physical object -- a
strong singularity. In the last several years, however, evidence has been
accumulating that the BKL singularity is not the only type of
singularity which may evolve in General Relativity from generic initial
data. 

The new type of singularity forms at the Cauchy horizon (CH) of spinning 
or charged black holes. (For a recent review see \cite{burko-ori-book}.) 
The features of this singularity are markedly
different from those of the BKL singularity: (i) It is {\it null} (rather
than spacelike). (ii) It is {\it weak} (according to Tipler's
classification
\cite{tipler-77}); Specifically, the tidal deformations which an
extended physical object suffers upon approaching the singularity are
bounded. In the case of a spinning black hole, 
the evidence for the null and weak singularity has emerged from analytical
perturbative \cite{ori-92,ori-97} and non-perturbative \cite{brady-98} 
analyses.
In addition, the  local existence and genericity of a null and weak
singularity in solutions of the vacuum Einstein equations was demonstrated
in \cite{ori-flanagan96}. This was more recently demonstrated also in the
framework of plane-symmetric spacetimes in \cite{ori-98}. For the toy
model of a spherical charged black hole, the main features of the CH
singularity were first found analytically for simplified models based on
null fluids \cite{hiscock-81,poisson-israel-90,ori-91}, and later
confirmed numerically for a model with a self-gravitating scalar field
\cite{brady-smith95,burko-97}. Expressions for the divergence rate
of the blue-shift factors for that model, which are valid everywhere along
the CH,
were found analytically in \cite{burko-ori98}. Those expressions are exact
on the CH as functions of retarded time. However, they are only asymptotic
expressions as functions of advanced time (see below).    

The strength of the null singularity is of crucial importance for the
question of the hypothetical possibility of hyperspace travel through the 
CH of black holes. A necessary condition for this possibility to be
realized is that
physical objects would traverse the CH peacefully. Because the CH is known
to be a curvature singularity, it is necessary that the singularity would
be weak according to the Tipler classification of singularity strengths. 
For the toy model of a spherical charged black hole, which we shall study
here, the properties of the CH singularity which have been found in 
\cite{poisson-israel-90,ori-91,brady-smith95,burko-97,burko-ori98} are all
consistent with the picture of a Tipler weak singularity. However, the
weakness of the singularity was demonstrated only for the simplified Ori
model \cite{ori-91} and at asymptotically early times for spinning black
holes \cite{ori-92}, where there are still no strong non-linear effects,
such as focusing of the null generators of the CH, which are crucial at
later times.  
In the context of spherical charged black holes and a self-gravitating
scalar field, several important features of the spacetime structure have
been found in fully non-linear numerical simulations. Specifically, it was
shown that for any point along the CH singularity 
there existed coordinates for which the metric coefficients were
finite and the metric determinant
was non-degenerate in an open neighborhood to the past 
\cite{brady-smith95,burko-97}.  
However, despite previous claims \cite{ori-92,burko-97,hod-piran98}, this
still does not guarantee that the singularity is weak in the Tipler sense
\cite{nolan-99}.

It is the purpose of this paper to present an analytical
demonstration of the weakness of the singularity for the model of a
spherical charged black hole with a self-gravitating, 
minimally-coupled, massless, real scalar field. Our results are valid at
arbitrary points
along the CH singularity, in particular at late times, where
strong non-linear effects (focusing of the null generators of the CH and
growth of the blue-shift factors) are crucial. In fact, our results are valid
everywhere along the CH singularity, down to the event where the
generators of the CH are completely focused, and the singularity becomes
spacelike and Tipler strong \cite{burko-99}. We emphasize that although
our discussion here is analytical, we do make assumptions which are based
on results obtained by numerical simulations. 

\section{Strength of the singualrity}
We write the general
spherically-symmetric line-element in the form 
\begin{equation}
\,ds^2=-f(u,v)\,du\,dv +r^2(u,v)\,d\Omega^2 ,
\label{metric}
\end{equation}
where $\,d\Omega^2=\,d\vartheta^2+\sin^2\vartheta\,d\varphi^2$ is the
line-element on the unit two-sphere. The coordinates $u,v$ are any
outgoing and ingoing null coordinates, correspondingly.  (Below, we shall
specialize to a specific choice of gauge, and define a particular choice 
of an ingoing null coordinate.)  
We consider the class of scalar field perturbations which is inherent to
any gravitational collapse process. These are the perturbations which
result from the evolution of non-vanishing multipole moments during the
collapse. When these perturbations propagate outwards, they are partially
reflected off the spacetime curvature and captured by the black hole.
This process results in a scalar field, which at late advanced times
decays along the event horizon like an inverse power of advanced time.     
Specifically, we assume that the
scalar field behaves along the event horizon at late times according to
$\Phi^{\rm EH}\propto (\kappa v_{e})^{-n}$
\cite{price-72,gundlach-94,burko-ori97}, where $n$ is a positive integer
which is related to the multipole moment under consideration. 
(We do not consider, however, other possible sources of perturbations 
\cite{burko-qed}.)  
By $v_{e}$ we denote the usual advanced time in the Eddington gauge, and
$r_{\pm},\kappa$ are the outer and inner horizons and the surface gravity 
of the latter, respectively, for a Reissner-Nordstr\"{o}m black hole
having the same external parameters as the black hole we consider has at
late times. 
We define the dimensionless ingoing Kruskal-like coordinate by 
$V\equiv -\exp(-\kappa v_{e})$. In the Kruskal gauge we denote the metric
function $g_{uV}$ by $-F/2$. 
For this model, it can be shown analytically that at arbitrary points
along the CH the following relations are satisfied, to the leading orders
in $\left[-\ln(-V)\right]^{-1}$ \cite{burko-ori98}:
\begin{eqnarray} 
r_{,V}&=&\frac{(nr_{-}A)^2}{rV}\left[-\ln\left(-V\right)\right]^{-2n-2}
\left\{1+ b_{1}\left[-\ln(-V)\right]^{-1} \right. 
\nonumber \\
&+&\left. b_{2}\left[-\ln(-V)\right]^{-2}
+O\left[-\ln(-V)\right]^{-3}
\right\}
\label{eq2}
\end{eqnarray}
\begin{eqnarray}
\Phi_{,V}&=&\frac{(nr_{-}A)}{rV}\left[-\ln\left(-V\right)\right]^{-n-1}
\left\{1+ c_{1}\left[-\ln(-V)\right]^{-1} \right.
\nonumber \\
&+&\left. c_{2}\left[-\ln(-V)\right]^{-2}
+O\left[-\ln(-V)\right]^{-3}
\right\}.
\label{eq3}
\end{eqnarray}
Here, $A=[r_+/(2r_-)](r_+/r_-+r_-/r_+)$, and the expansion
coefficients $b_i$ and $c_i$ are functions
of retarded time only.  Note that in the limit $V\to 0$ these are exact
expressions as functions of retarded time. That is, to the leading order
in $\left[-\ln(-V)\right]^{-1}$ there is implicit dependence on retarded
time through $r=r(u)$, and along the CH singularity both $r_{,V}$ and 
$\Phi_{,V}$ are exactly inversely proportional to $r(u)$, in the following
sense. Consider two outgoing null rays, and let one ray be at $u=u_1$,
say, and the other at $u=u_2$. The ratios $r_{,V}(2)/r_{,V}(1)$ and 
$\Phi_{,V}(2)/\Phi_{,V}(1)$ approach $r(u_1)/r(u_2)$ as $V\to 0$. Taking
now $u_1$ to be in the asymptotically-early parts of the CH, where
$r(u_1)\approx r_{-}$, we find that both $r_{,V}$ and $\Phi_{,V}$ are
inversely proportional to $r(u)$. As $r(u)$ is 
monotonically decreasing as a function of retarded time along the CH, we
find that $r_{,V}$ and $\Phi_{,V}$ grow monotonically along the CH. This
growth is a non-linear effect which indicates the strengthening of the
singularity along the CH (although the singularity is still weak according
to the Tipler classification; see below).  

All the non-zero components of
the Riemann-Christoffel curvature tensor $R_{\mu\nu\rho\sigma}$ 
are given completely in terms of the divergent blue-shift factors  
$r_{,V},\Phi_{,V}$, and the {\it finite} quantities $r,r_{,u},\Phi_{,u}$,
and $F$. Interestingly, $R_{\mu\nu\rho\sigma}$ does not
depend on gradients of $F$. This can be understood from the following
consideration. The tensor $R_{\mu\nu\rho\sigma}$ can be written
as the sum of the Weyl tensor, and another tensor which is built from the
Ricci and the metric tensors (but not involving their derivatives). In
spherical symmetry the Weyl tensor is given completely in
terms of the mass function, which is defined by 
$r_{,\mu}r^{,\mu}=1-2m(u,v)/r+q^2/r^2$, $q$ being the charge of the black
hole. In Kruskal-like coordinates the mass function 
$m(u,V)=(r/2)(1+4r_{,u}r_{,V}/F)+q^2/(2r)$,  
which depends only on $r,r_{,V},r_{,u}$, and
$F$. (The divergence of the mass function at the singularity, and
consequently also the divergence of curvature, is evident from the
divergence of $r_{,V}$ and the finiteness of $r$, $r_{,u}$ ,and $F$.) 
The Ricci tensor $R_{\mu\nu}=2\Phi_{,\mu}\Phi_{,\nu}$, and
consequently $R_{\mu\nu\rho\sigma}$ is independent of gradients of $F$.

We find that the components of $R_{\mu\nu\rho\sigma}$ which have the
strongest divergence near the CH are 
$R_{V\varphi V\varphi}$ and $R_{V\vartheta V\vartheta}$. (In the Appendix
we list all the non-zero independent components of the Riemann-Christoffel
tensor.) 
It can be readily shown that 
\begin{equation}
R_{V\vartheta V\vartheta}
=-r^2\left(\Phi_{,V}\right)^2=\,\sin^{-2}\vartheta\,R_{V\varphi V\varphi}.
\end{equation} 
We denote these two components 
schematically and collectively by ${\cal R}$. 
We find that, to the leading orders in
$V$ and in $\left[-\ln\left(-V\right)\right]^{-1}$, 
\begin{equation}
{\cal R}(V)\propto V^{-2}\left[-\ln\left(-V\right)\right]^{-2n-2}.
\end{equation}
(The other divergent components of $R_{\mu\nu\rho\sigma}$, i.e., 
$R_{uVuV}, R_{\vartheta\varphi\vartheta\varphi}, 
R_{u\vartheta V\vartheta}$, and $R_{u\varphi V\varphi}$ are proportional
to the leading order in $V$ to $V^{-1}$ times a logarithmic factor.)
Because both metric functions $r$ and $F$ have finite values at the CH
(which is known from the numerical simulations of
\cite{brady-smith95,burko-97}), 
it is easy to show that the dependence of ${\cal R}(V)$ on $V$ does not
change when we transform to a parallelly-propagated frame. 

We next find
$V(\tau)$ as a function of affine parameter (proper time) $\tau$ along a
general null (timelike) geodesic. For general causal geodesics, the
geodesic
equations are
\begin{equation}
\dot{u}\dot{v}=\left(mr^2\dot{\Omega}^2-p\right)
\label{ge1}
\end{equation}
\begin{equation}
\ddot{v}+(f_{,v}/f)\dot{v}^2+2mr_{,u}\left(f\dot{u}\dot{v}+p\right)/(rf)
=0
\label{ge2}
\end{equation}
\begin{equation}
\ddot{u}+(f_{,u}/f)\dot{u}^2+2mr_{,v}\left(f\dot{u}\dot{v}+p\right)/(rf)
=0.
\label{ge3}
\end{equation}
Here, $m=0(1)$ for radial (non-radial) geodesics, and $p=0(-1)$ for null
(timelike) geodesics. A dot denotes differentiation with respect to affine
parameter (proper time), and $\dot{\Omega}^2=\dot{\vartheta}^2+
\,\sin^2\vartheta\,\dot{\varphi}^2$. The geodesic equations can be solved
to the leading order in $[\ln (-V)]^{-1}$ for all causal geodesics. This
is done by using the field equation \cite{burko-97}
\begin{equation}
F_{,V}/F=r_{,VV}/r_{,V}+r(\Phi_{,V})^2/r_{,V}
\label{field}
\end{equation}
to find $F_{,V}/F$ explicitly. Substituting Eqs. (\ref{eq2}) and
(\ref{eq3}) in Eq. (\ref{field}) we find 
\begin{eqnarray}
\left(\ln F\right)_{,V}&=&\left[\ln\left(-r_{,V}\right)\right]_{,V}
+\frac{1}{V}+\left(2c_1-b_1\right)\frac{1}{V}\left[-\ln\left(-V\right)
\right]^{-1}\nonumber \\
&+&\left(b_1^2-b_2-2b_1c_1+c_1^2+2c_2\right)\frac{1}{V}
\left[-\ln\left(-V\right)\right]^{-2}\nonumber \\
&+&O\left\{
\frac{1}{V}\left[-\ln\left(-V\right)
\right]^{-3}\right\}.
\end{eqnarray}
Integration yields
\begin{eqnarray}
\ln F&=&\ln\tilde{F}_0+\ln\left(Vr_{,V}\right)-
\left(2c_1-b_1\right)\left\{\ln\left[-\ln\left(-V\right)\right]\right\}
\nonumber \\
&+&\left(b_1^2-b_2-2b_1c_1+c_1^2+2c_2\right)
\left[-\ln\left(-V\right)\right]^{-1}\nonumber \\
&+&O\left\{\left[-\ln\left(-V\right)\right]^{-2}\right\}.
\end{eqnarray}
Here, $\ln\tilde{F}_0$ is an integration constant, which can be a
function of $u$. Exponentiating both
sides, and substituting Eq. (\ref{eq2}) for $r_{,V}$ we find 
\begin{eqnarray}
F&=&\tilde{F}_0\frac{\left(nr_-A\right)^2}{r}
\left[-\ln\left(-V\right)\right]^{-2n-2+b_1-2c_1}\nonumber \\
&\times &\left\{1+b_1\left[-\ln\left(-V\right)\right]^{-1}
+O\left[-\ln\left(-V\right)\right]^{-2}\right\}\nonumber \\
&\times &\exp\left\{\left(b_1^2-b_2-2b_1c_1+c_1^2+2c_2\right)
\left[-\ln\left(-V\right)\right]^{-1}\right.\nonumber \\
&+& \left. O\left[-\ln\left(-V\right)\right]^{-2}\right\}.
\end{eqnarray}
From the numerical results of \cite{brady-smith95,burko-97} it is known
that as $V\to 0$, $F$ approaches a finite value. Consequently, in order to
have consistency with the numerical results we require that 
$b_{1}-2c_{1}=2n+2$, which implies that 
\begin{eqnarray}
F=F_0(u)
\left\{1+B\left[-\ln\left(-V\right)\right]^{-1} 
+ O\left[-\ln\left(-V\right)\right]^{-2}\right\},
\end{eqnarray}
where $B=(2n+3)b_1-b_2+c_1^2+2c_2$, and $F_0(u)=\tilde{F}_0(nr_-A)^2/r$. 
For the logarithmic derivative of $F$ we find, to the leading order in 
$\left[-\ln\left(-V\right)\right]^{-1}$, that  
\begin{equation}
F_{,V}/F=
B\frac{1}{V}
\left[-\ln\left(-V\right)\right]^{-2}. 
\label{log-der}
\end{equation}
(Higher order terms in $\left[-\ln\left(-V\right)\right]^{-1}$ are
functions of retarded time.) 
We note that only $b_1$ and $c_1$ are constrained. The coefficients of
highr-order terms in $\left[-\ln\left(-V\right)\right]^{-1}$ are
immaterial near the CH for our determination of the strength of the
singularity. 
Note that $F\to F_0(u)$ as $V\to 0$, and that $F$
is not analytic in $V$. 
In fact, this is an important property of the CH singularity: 
In a kruskal-like gauge the metric
functions $r$ and $F$ are finite at the singularity, but their gradients
in the outgoing direction diverge. The finiteness of $r$ and $F$ at the CH
also implies that the metric determinant is non-degenerate. 
This expression for $F$ is similar to
the behavior of the $g_{uV}$ metric function found for the simplified Ori
model \cite{ori-91}. 
We stress that although this expression for $F$ is exactly
valid everywhere along the CH singularity, it still does not allow us to
find the variation of $F$ with retarded time, as we do not know
the form of $F_0(u)$ or $r(u)$ along the CH. We note that near the CH
the metric function $F$ is monotonic in $V$. This result is in accord with
the numerical results of \cite{burko-97}. (Notice, however, the
disagreement with the numerical results of \cite{hod-piran98}. It is
reasonable to expect the behavior of the metric functions near the CH to
be similar for both cases of real and complex scalar fields. The lower
panels of Figs. (3) of \cite{hod-piran98} imply, however, a non-monotonic 
behavior of $F$. That kind of behavior can be obtained from a numerical
code with a specific choice of parameters if the latter is far from
convergence near the CH.)

Let us consider first radial geodesics. (The case of non-radial geodesics
will be treated next.) In the null case ($m=0$ and $p=0$) it is
easy to solve the geodesic equations (\ref{ge1})--(\ref{ge3}).  For
outgoing geodesics one readily finds that the solution is 
$u={\rm const}$ and $\dot{V}={\rm const}/F$. The metric function $F$ can
be expanded in $[-\ln (-V)]^{-1}$, despite its non-analyticity in $V$. To
the leading orders in $[-\ln (-V)]^{-1}$ we find that 
\begin{equation}
F=F_0(u) 
\left\{1+B\left[-\ln\left(-V\right)\right]^{-1}\right\},
\end{equation}
such that 
\begin{equation}
\dot{V}=\frac{\rm 
const}{F_0}\left\{1-B\left[-\ln\left(-V\right)\right]^{-1}\right\}.
\end{equation}
The solution for
$V(\tau)$ is then given asymptotically close to the CH, to the leading
orders in $[-\ln (-V)]^{-1}$ by  
\begin{equation}
V(\tau)=\tau\left\{1-B
\left[-\ln\left(-\tau\right)\right]^{-1}\right\}.
\end{equation}
(Recall that the affine parameter is given up to a linear transformation.) 
To the leading order in 
$\left[-\ln\left(-\tau\right)\right]^{-1}$ we can thus approximate
$V(\tau)\approx\tau$. 
Note that although asymptotically $V(\tau)$ and $\dot{V}(\tau)$ behave
like $\tau$ and $\dot{\tau}\equiv 1$, correspondingly, $\ddot{V}(\tau)$
behaves very differently from $\ddot{\tau}\equiv 0$. In fact, 
$\ddot{V}(\tau)$ diverges as $\tau \to 0$. 
Therefore, one can approximate $V(\tau)\approx\tau$ only if one is
interested in $V(\tau)$ itself, or at the most in $\dot{V}(\tau)$. This
approximation is invalid for $\ddot{V}(\tau)$ or higher derivatives. 
For radial timelike geodesics ($m=0$ and $p=-1$) one uses the finiteness
of $F_{,u}$ to find approximately that again 
$\dot{V}\approx {\rm const}/F$, and consequently one finds the same result
for 
$V(\tau)$. In the case of non-radial geodesics ($m=1$) one can consider a
specific value of the retarded time at which the geodesic hits the CH
singularity. Than, $r,r_{,u}$, and $F$ can be approximated by their values
at the singularity. When this is done, the equations for null non-radial 
($m=1$ and $p=0$) geodesics become inhomogeneous linear equations. The 
corresponding homogeneous equations are nothing but the equations for the
radial geodesics, which we already solved. Particular solutions for the
inhomogeneous equations are easy to generate, and one finds that again
$V(\tau)$ is given asymptotically as before. The last case is the case of
non-radial
timelike 
($m=1$ and $p=-1$) geodesics. In this case the geodesic equation becomes 
(under similar assumptions) an inhomogeneous non-linear equation. Although
this equation is hard to solve directly, it can be checked that the same
leading order proportionality of $V(\tau)$ and $\tau$ is the solution
also for this case. 
We thus find that
for all causal geodesics, to the leading order in
$\left[-\ln\left(-\tau\right)\right]^{-1}$, $V(\tau)$ is proportional to
$\tau$. 

We next re-express ${\cal R}$ as a function of affine parameter (proper
time) along radial or non-radial null (timelike) geodesics. To the
leading order in $\left[-\ln\left(-\tau\right)\right]^{-1}$ we find that
in a parallelly-propagated frame 
\begin{equation}
{\cal R}(\tau)\propto 
\tau^{-2}\left[-\ln\left(-\tau\right)\right]^{-2n-2}.
\end{equation}
A necessary condition for a singularity to be strong in the Tipler sense
is given by the following theorem \cite{clarke-krolak85}: For null
(timelike) geodesics, if the singularity is strong in the Tipler sense,
then for at least one component of the Riemann-Christoffel curvature
tensor in a parallelly-propagated frame, the twice integrated component
with respect to affine parameter (proper time) does not converge at the
singularity. Specifically, the necessary condition for the singularity to
be Tipler strong is that 
\begin{equation}
{\cal I}(\tau)=\int^{\tau}\,d\tau '\int^{\tau '}\,d\tau ''
\left|{\cal R}(\tau '')\right|
\end{equation}
does not converge as $\tau\to 0$. 
It can be readily shown that when ${\cal R}(\tau)$ is integrated twice
with respect to $\tau$,
${\cal I}(\tau)$ converges in the limit $\tau\to 0$. Consequently, we find
that a necessary condition for any causal geodesic to terminate at a
Tipler strong singularity is not satisfied. Hence, all causal geodesics
terminate at a Tipler weak singularity, namely, the singularity is
Tipler weak. The physical content of this result is that the volume
element of physical objects remains bounded at the singularity. We
emphasize that this result is valid everywhere along the singularity, in
particular at late retarded times where the non-linear effects (focusing
of the
generators of the CH and the growth of the blue-shift factors) are 
crucial.

\section{Concluding remarks}

Recently, Ori suggested to define a deformationally-strong singularity in
the following way. Let $\lambda (\tau)$ be a timelike geodesic with proper
time $\tau$ along it.  
The geodesic $\lambda (\tau)$
terminates at a deformationally-strong singularity at $\tau =0$ if at
least one of the following two conditions holds: 
(i) $\lambda (\tau)$
terminates at a Tipler strong singularity, or (ii) there exists a Jacobi
field ${\bf J}(\tau )$ for which at least one parallelly-propagated tetrad
component is unbounded at the limit $\tau\to 0$ \cite{ori-private}. 
This definition is more physically-motivated than Tipler's definition,
because it classifies a singularity as strong not only when the volume
element vanishes, but also when the volume element diverges to infinity,
or there is infinite compression in one direction, and infinite stretching
in a different direction, such that the volume element remains bounded.  
In fact, it can be shown that the failure of the necessary condition for
the singularity to be Tipler strong implies 
not only the boundedness of the volume element, 
but also the boundedness of the Jacobi fields themselves 
\cite{ori-private}, such that objects are not expected to be destroyed 
also because of distortions which preserve the volume element or
divergence to infinity of the volume element.
Consequently, the singularity we are studying here 
is weak also in the sense of Ori (deformationally weak). (Ori's definition
does not include null geodesics. However, extended physical objects move
along timelike geodesics, such that this deficiency 
does not restrict our discussion. It is conceivable that both Ori's
definition and theorem for the necessary condition for the singularity to
be Ori strong can be generalized to all causal geodesics.) 

We note that according to the Kr\'{o}lak classification of
singularities \cite{krolak-87} 
this is a strong singularity. Specifically, if we
integrate over the divergent components of the Riemann-Christoffel tensor
only once, the integral does not converge on the singularity. This means
that the expansion diverges (negatively) on the singularity
(Kr\'{o}lak strong), but still the volume element 
(and the distortion in general) remains finite (Tipler and Ori weak). One
might be worried that even if spacetime were classically extendible beyond
the CH, this infinite negative expansion would inevitably result in
unavoidable destruction of any extended physical object subsequent to its
traversing of the CH \cite{herman-hiscock92}. 
Of course, any classical extension of
geometry beyond the CH is not unique. We can, however, consider an
extension with a continuous ($C^0$) metric and a {\it unique} $C^1$
timelike geodesic, and assume that the object follows this geodesic
\cite{ori-book}. Any extension of classical
geometry beyond the CH (which can be modeled as a thin layer wherein the
geometry is inherently quantum) requires an infinite flux of
{\it negative} energy traveling along the contracting CH. This negative
energy flux may then act to regularize the expansion, such that the
deformation rate of physical objects beyond the CH would be bounded
\cite{ori-book}. (The infinite expansion is likely not to destroy physical
objects up to the CH \cite{ori-book}, in contrast with Ref. 
\cite{herman-hiscock92}.)  
Indeed, a simplified two-dimensional quantum model shows
an infinite ingoing flux of {\it negative} energy along the CH
\cite{hiscock-77,birrell-davis78}. More recent semi-classical toy models
of a quantum field on a mass-inflation background are not inconsistent
with this picture \cite{balbinot-poisson93,anderson-93}. One should not
take these quantum results too seriously, however, because in these models
the semi-classical contributions are dominated by the regime where
curvature is Planckian, such that the semi-classical approximation
is not expected to be valid anymore. Instead, a full quantum theory of
gravity is of need. 
Of course, in the absence of a valid
theory of quantum gravity it is difficult to make predictions on the
detailed interaction of the thin layer of the CH with physical objects,
but the evidence we currently have do not preclude the possibility of
objects traversing the CH singularity peacefully.

I have benefited from useful discussions with Patrick Brady 
and Amos Ori. This work was supported by NSF grant
AST-9731698.

\section*{Appendix}
The independent components of the Riemann-Christoffel curvature
tensor which do not vanish identically are:

\begin{eqnarray*}
R_{uvuv}&=&-\frac{1}{4}\frac{f}{r^2}\left[4r_{,u}r_{,v}+f\left(1
-2\frac{q^2}{r^2}\right)\right]+f\Phi_{,u}\Phi_{,v}\\
R_{\vartheta\varphi\vartheta\varphi}&=&4\frac{r^2}{f}
r_{,u}r_{,v}\; \sin^2\vartheta\\
R_{u\vartheta
v\vartheta}&=&-r_{,u}r_{,v}-\frac{f}{4}\left(1-\frac{q^2}{r^2}\right)\\
R_{v\vartheta v\vartheta}&=& -r^2\Phi_{,v}^2\\
R_{u\varphi v\varphi}&=&
-\left[r_{,u}r_{,v}+\frac{f}{4}\left(1-\frac{q^2}{r^2}\right)\right]
\;\sin^2\vartheta \\
R_{u\varphi u\varphi}&=&-r^2\Phi_{,u}^2\;\sin^2\vartheta \\
R_{u\vartheta u\vartheta}&=&-r^2\Phi_{,u}^2 \\
R_{v\varphi v\varphi}&=&-r^2\Phi_{,v}^2\;\sin^2\vartheta
\end{eqnarray*}

\end{document}